\documentclass[%
 reprint,
 amsmath,amssymb,
 aps,
pra,
]{revtex4-1}
\usepackage{graphicx}
\usepackage{dcolumn}
\usepackage{bm}
\usepackage{graphicx}
\usepackage{multirow}
\usepackage{makecell}

\usepackage[dvipdfm,colorlinks,linkcolor=blue, urlcolor=blue, anchorcolor=blue, citecolor=blue]{hyperref}
\begin{document}

\preprint{APS/123-QED}

\title{Controllable multiple beam splitting in Hermitian and non-Hermitian symmetric coupled waveguide systems}

\author{Fu-Quan Dou}
\email{doufq@nwnu.edu.cn}
\affiliation{College of Physics and Electronic Engineering, Northwest Normal University, Lanzhou, 730070, China}
\author{Ya-Ting Wei}
\affiliation{College of Physics and Electronic Engineering, Northwest Normal University, Lanzhou, 730070, China}
\author{Min-Peng Han}
\affiliation{College of Physics and Electronic Engineering, Northwest Normal University, Lanzhou, 730070, China}
\author{Jian-An Sun}
\email{sunja@nwnu.edu.cn}
\affiliation{College of Physics and Electronic Engineering, Northwest Normal University, Lanzhou, 730070, China}

\begin{abstract}
We investigate high-fidelity multiple beam splitting in Hermitian and non-Hermitian symmetric coupled waveguides (WGs) with one input and $2N$ output waveguide channels. In Hermitian systems, we realize adiabatically light splitting in resonant case based on the stimulated Raman adiabatic passage (STIRAP) and arbitrary proportion from the middle WG to outer WGs in propagation coefficients  mismatch case using shortcuts to adiabaticity (STA) technique. In non-Hermitian systems with even WGs being dissipative, the compact and robust beam splitting can be achieved by eliminating the non-adiabatic coupling via the non-Hermitian STA method. We further verify the feasibility of our theoretical predictions by means of the beam propagation method (BPM). The suggested multiple beam splitters open new opportunities for the realization of on-chip high-bandwidth photonics with high fidelity in short distances.
\end{abstract}

\maketitle

\section{\label{sec:level1}Introduction}
The beam splitter is an essiential element in integrated optics which has attracted considerable attention for a wide range of applications such as power transfer between WGs, quantum communication, magneto-optic data storage and many other practical fields \cite{180591,Sun:09,PhysRevLett.101.200502,Watts:05}. Beam splitters are devices in which the optical signal is input in one WG and then output to two or more WGs with the same or different intensities. To our knowledge the simplest and most advantageous is $1\times 2$ beam splitter created by Y-branch \cite{PhysRevLett.82.2564,Frandsen:04} and T-branch junctions \cite{Fan:01,Boscolo:02,Hong2018}. However, mainly motivated by the development of on-chip high-bandwidth photonics which requires a larger number of output ports. The multiway beam splitters form the basic building blocks in wavelength division multiplexing systems and in fibre to home networks \cite{Yu2008}, and are critically essential for the implementation of quantum logic gates in future quantum computers \cite{O'Brien2003,Li2021}. There has been a growing trend in the realization of multiple WGs ($1\times N$) \cite{Yu2008,5290043,5702357,PhysRevA.85.055803,Tseng:10,Chung:12,Shi2016,Amal2016,Franz2021,Ciret:12,Lunghi:18,Sie:19,PhysRevA.88.013808} or from multiple to multiple channels ($N\times N$) \cite{Leuthold2001}.

The analogy between quantum mechanics and WG optics has led to many protocols based on quantum techniques to manipulate light in WG arrays \cite{PhysRevA.71.065801,PhysRevLett.101.193901,PhysRevA.82.043818,PhysRevA.97.023811}. Specifically, the one related to quantum process of stimulated Raman adiabatic passage (STIRAP)  \cite{PhysRevA.82.043818,PhysRevA.97.023811,PhysRevA.70.063409,PhysRevA.71.065801,PhysRevLett.101.193901,Longhi2009,RevModPhys.89.015006} is a powerful tool for achieving complete and stable power transfer in WG optics. Various types of STIRAP-like optical devices have been presented, and hence they could enjoy the same advantages as STIRAP in terms of efficiency and robustness against variations of the experimental parameters, such as the WGs couplings, the distance between the WGs and their geometry \cite{PhysRevA.93.033802,PhysRevA.87.013806,Ciret:12,Dou2020164516}. On the contrary, conventional integrated beam splitters (for instance based on resonant coupling) are designed for the energy that changes periodically. Only those at a specific position can obtain the perfect energy output. Despite the advantages, adiabatic beam-splitting usually requires a sufficiently long distance or large coupling constant or both, which are undesirable in many experimental applications. To speed up the adiabatic process and design higher fidelity devices \cite{PhysRevA.89.012123,PhysRevA.93.043419,Dou2020164516,PhysRevA.98.022102,Dou2016,Aashna_2017}, shortcut to adiabaticity (STA) was employed to the coupled WG systems \cite{Tseng:14,Ho:15,PhysRevA.79.055802,Huang2019,Martinez-Garaot:17,Chen:16}.

In practice, when considering the realistic propagation of light in the WG, the propagation loss may be unavoidable \cite{Olivier:03,PhysRevB.80.035123,PhysRevA.84.023415,PhysRevA.103.023527}. This dissipative effect has, in many case, a negative impact on shaping the optical beams. However, it also causes a problem about the stability of such WG structures \cite{OFaolain:10,Wu:16,Chung:12,PhysRevA.91.023822}. Recent work has also pointed out that a simple model of three parallel WGs with the central one being dissipative can exhibit an ultra-broadband power splitting associated with an overall half power loss \cite{PhysRevA.103.023527}. It is worth noting that when taking account of external control parameters such as gain or loss ( i.e., imaginary refractive index) or both, one will obtain a non-Hermitian Hamiltonian \cite{PhysRevA.89.063412,PhysRevA.91.023822,PhysRevA.99.063834,Chen2018} for the description of the WG system. Models based on non-Hermitian Hamiltonian could offer a range of surprising and potentially useful phenomena for shaping the optical beams \cite{PhysRevA.87.052502,Ke2018,Savoia2016} in contrast to traditional conservative or low-loss structures. Especially in the context of parity-time symmetric systems \cite{El-Ganainy:07,Zhao2018,El-Ganainy2018}, they could produce a faster evolution than a Hermitian one while keeping the eigen energy difference fixed \cite{Uzdin_2012}. Moreover, drawing inspiration from STA to Hermitian situations, it has been successfully extended to non-Hermitian Hamiltonian as well \cite{PhysRevA.87.052502,PhysRevA.89.063412,Li2017,PhysRevA.89.033403,Zhang_2021,Li:17,tang2020hermitian,li2019quantum}.

In this paper, we propose adiabatic optical beam splitters with one input and $2N$ $(N\geq2)$ output WGs and observe the evolution of the system both for the Hermitian and non-Hermitian cases. We could achieve the adiabatic light splitting in resonant case based on STIRAP and the arbitrary proportion from the middle WGs to outer WGs in propagation coefficients mismatch case using STA technique. In addition, the compact and robust beam splitting is also achieved via the non-Hermitian STA method in non-Hermitian systems with the even WGs being dissipative. Furthermore, we verify the feasibility of our theoretical predictions by using beam propagation method (BPM) \cite{kawano2004introduction}.

This paper is organized as follows. In Sec. \ref{sec:level2}, we introduce the multiple beam splitting optical model and its Hamiltonian of symmetric coupled WGs. In Sec. \ref{sec:level3}, we firstly present a simplest case, where there is no detuning between these WGs and dissipation. We apply STIRAP to design $1\times2 $ beam splitter. Secondly, we consider the phase mismatch coupling model with which we can achieve arbitrary proportional beam splitting via STA. Then we consider the effects of the detuning and dissipation on the conversion process and further design short and robust $1\times2$ beam splitters by the non-Hermitian STA. In adddition, we investigate the fabrication tolerances of our beam splitters and compare the parameters and performance of our works with others. Numerical calculations are also verified by optical simulations based on BPM in Sec. \ref{sec:level4}. Finally, the conclusions are given in Sec. \ref{sec:level5}.
\section{\label{sec:level2} Model and Hamiltonian}
\begin{figure}[htbp]
 \centering
 \includegraphics[width=0.4\textwidth]{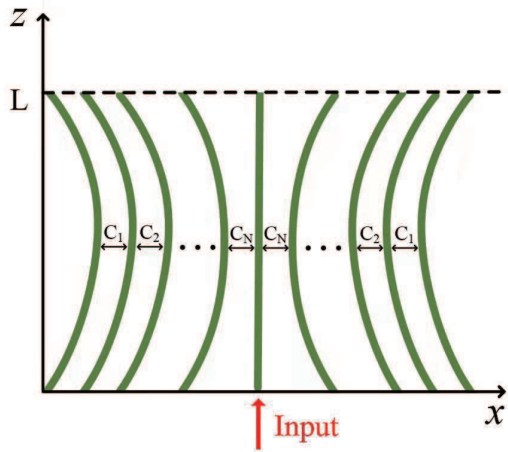}
 \caption{Schematic of the WG structure for multiple splitting of light. Here a Gaussian-shaped light beam is injected initially in the central WG $N+1$.}
\label{fig1}
\end{figure}

The WG structure leading to the multiple beam splitting is shown in Fig. \ref{fig1}. Output WGs are characterized by some specific curved structures, and the middle WG is in straight structure. For a system of one input and $2N$ output WGs, the evolution of the wave amplitude with nearest-neighbor evanescent coupling is accurately described by a set of $2N+1$ coupled differential equations (in matrix form),
\begin{equation}
\label{eq1}
i\frac{d}{d z}\mathbf{a}(z)=H(z)\mathbf{a}(z),
\end{equation}
where $\mathbf{a}(z)=[a_{1}(z),a_{2}(z),\ldots, a_{2N+1}(z)]^T$ is the light amplitude in the individual WGs and the Hamiltonian in the rotating-wave approximation reads as
\begin{equation}
\setlength{\arraycolsep}{2.5pt}
H=\left[\begin{array}{ccccccc}
     0 & C_{1} & 0 & 0 & 0 &\cdots&0\\
     C_{1} & \Delta-i\Gamma & \ddots&0&0 &\cdots&0\\
     0 & \ddots & \ddots & C_{N} & 0&\cdots&0\\
     0 & \cdots &C_{N}&\Delta-i\Gamma&C_{N}&\cdots&0\\
     0 &\cdots & 0 & C_{N} & \ddots &\ddots&0\\
     0&\cdots &0&0 & \ddots & \Delta-i\Gamma & C_{1}\\
     0&\cdots &0&0 & 0 & C_{1} & 0
\end{array}\right],
\end{equation}
where the $C_{k}$ $(k=1,2,3\cdots N)$ is the $z$-dependent coupling coefficient corresponding to the physical parameters of WGs and the WG spacing. In our cases, the symmetrical form of WGs insures that, $C_{k}$ is the coupling coefficients of the $k-$th to $(k+1)-$th WG, and of the $(2N+1-k)$-th to $(2N+2-k)$-th WG. We assume the odd WGs are identical and lossless, while the even WGs are dissipative. The related loss is denoted by the constant $\Gamma$ which allows a relatively easy implementation of the non-Hermitian Hamiltonian in WG optics. This non-Hermitian term can also be realized conveniently in the experiments. Recent experiment has proposed the loss rate can be controlled by the number of Fe$^{2+}$ in LiNbO$_{3}$ optical WGs \cite{Ruter:10}. Nevertheless, the even WGs may also differ from the other WGs in terms of geometrical dimensions and refractive index contrast, leading to the propagation constant mismatch $\Delta$ which is linearly dependent on the width difference. In addition, at the central position of the structure, the adjacent WGs have minimum distances and correspond to maximum couplings. Theoretically, the WG length in Fig. \ref{fig1} are supposed to be infinitely long. Therefore, the couplings at the inputs and outputs of the WGs can be safely ignored \cite{PhysRevA.87.013806}.
\section{\label{sec:level3} Controllable multiple beam splitting for array of WGs}
For the sake of simplicity, we now present an example for the splitting of light in adiabatic Hermitian and non-Hermitian five-WG systems. In the non-Hermitian case, the two outermost WGs (WG1 and WG5) and the central one (WG3) are supposed to be identical and lossless, while the two outer WGs (WG2 and WG4) are dissipative. As we mentioned above, we are interested in the case where the light is initially launched in the middle WG (WG3), and then export to the outer WGs with the Hamiltonian given by
\begin{equation}
H(z)=\left[\begin{array}{ccccc}
     0 & C_{1} & 0 & 0 & 0 \\
     C_{1} & \Delta-i\Gamma & C_{2} & 0 & 0 \\
     0 & C_{2} & 0 & C_{2} & 0\\
     0 & 0 & C_{2} & \Delta-i\Gamma & C_{1} \\
     0 & 0 & 0 & C_{1} & 0
\end{array}\right].
\end{equation}
As we all know, it is a quite helpful treatment to predict the dynamics in optical WGs by analyzing the eigenstates of the Hamiltonian. For this reason, in the following three cases, we will calculate the eigenstates of corresponding Hamiltonian to discuss the evolution of light in optical WGs. In all the cases, we consider Gaussian coupling coefficients $C_{1}$ and $C_{2}$,
\begin{eqnarray}
\label{eq5}
C_{1,2}&=C_{0}\exp \lbrack -\alpha(z-z_{1,2})^2\rbrack.
\end{eqnarray}
\subsection{Case 1: $\Delta=0,\Gamma=0$}
In the first example, we consider the simplest case that the WGs are resonant ($\Delta=0$) and lossless ($\Gamma=0$). Thus, we can quantify the original Hamiltonian and its corresponding eigenvalues

\begin{equation}
\begin{aligned}
E_{1}&=0,\\
E_{2,4}&=\pm C_{1},\\
E_{3,5}&=\pm\sqrt{C^2_{1}+ 2C^2_{2}}.\\
\end{aligned}
\end{equation}
The quantities $C_{1}$ and $C_{2}$ are similar to the Rabi frequencies for the Pump and Stokes pulses in the STIRAP process in quantum physics. For an odd number of WGs in the array, the eigenvector corresponding to the zero eigenvalue is the so-called dark state and we choose this state as our instantaneous eigenstate:
\begin{eqnarray}
|\Phi_{1}(z)\rangle=\frac{\sqrt{2}}{2}\sin(\theta)|1\rangle-\cos(\theta)|3\rangle+\frac{\sqrt{2}}{2}\sin(\theta)|5\rangle,
\end{eqnarray}
where the mixing angle $\theta = \theta(z)$ is defined as
\begin{eqnarray}
\tan(\theta)=\frac{2\sqrt{C_{2}}}{C_{1}}.
\end{eqnarray}
The adiabatic Hamiltonian can be derived as
\begin{equation}
\setlength{\arraycolsep}{1.1pt}
H_{a}(z)=\left[\begin{array}{ccccc}
     0 & 0 & -\frac{i\dot{\theta}}{\sqrt{2}} & 0 & 0 \\
     0 & C_{1} & 0 & 0 & 0 \\
     \frac{i\dot{\theta}}{\sqrt{2}} & 0 & \sqrt{C^2_{1}+ 2C^2_{2}} & 0 & \frac{i\dot{\theta}}{\sqrt{2}} \\
     0 & 0 & 0 & -C_{1} & 0 \\
     0 & 0 & -\frac{i\dot{\theta}}{\sqrt{2}} & 0 & -\sqrt{C^2_{1}+ 2C^2_{2}}
\end{array}\right].
\end{equation}

In the normal adiabatic evolution (i.e., STIRAP), there are no transitions between adiabatic states when the adiabatical conditions is well fulfilled. That is, the difference between the diagonal elements of $H_{a}(z)$ must be much larger than the off-diagonal elements. This occurs when
\begin{eqnarray}
|\dot{\theta}|&\ll|C_{1}|, |\dot{\theta}|&\ll|\sqrt{C^2_{1}+ 2C^2_{2}} |,
\label{eq9}
\end{eqnarray}
which requires that $C_{1}$ and $C_{2}$ vary smoothly with $z$. We note that in relation to coupled WG devices the adiabatic condition entails long device length for the realization of high fidelity adiabatic light transfer.
\begin{figure}[htbp]
\centering
\includegraphics[width=0.45\textwidth]{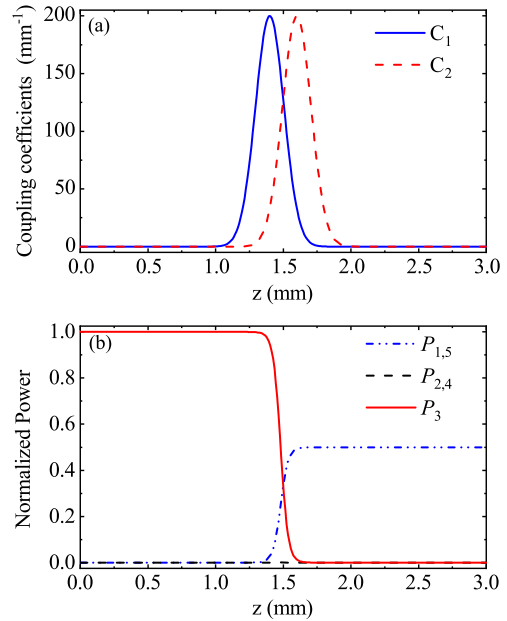}
\caption{(a) Coupling coefficients $C_{1}$ and $C_{2}$ with shapes as in Eq. (\ref{eq5}). (b) Spatial evolution of the powers $P_{3}$ (solid red lines), $P_{1,5}$ (double dotted blue lines), and $P_{2,4}$ (dashed black lines) in the five WGs upon injection in WG3. Parameters: $C_{0} =200 $mm$^{-1}$, $\alpha=45$, $z_{1}=1.4$ mm and $z_{2}=1.6$ mm.}
\label{fig2}
\end{figure}
We assume the input as $\mathbf{a}(z_{i})=\lbrack 0,0,1,0,0\rbrack^T$, for the following initial and final conditions:
\begin{eqnarray}
\label{eq10}
\theta_{i}=0,  \theta_{f}=\frac{\pi}{2},
\end{eqnarray}
which leads to the following final light field distributions among WGs: $\mathbf{a}(z_{f})=\lbrack \frac{\sqrt{2}}{2},0,0,0,\frac{\sqrt{2}}{2}\rbrack^T$ as shown in Fig. \ref{fig2}. When the adiabatic condition is satisfied, the change of the variable coupling coefficients is slow, a one-to-two beam splitting can be achieved, from the middle WG3 to the outermost WG1 and WG5 with the same intensity ($P_{1,5}=0.5$)  after propagation, as described by the dark state $|\Phi_{1}(z)\rangle$, a coherent superposition of states $|1\rangle$ and $|5\rangle$ can be created, so the outer WG2 and WG4 are almost unpopulated during evolution. Indeed, we can also manipulate the ratio of beam splitting at will depending on the single parameter $\theta$.

\subsection{Case 2: $\Delta\neq 0,\Gamma=0$}
\begin{figure*}[htbp]
\centering
\includegraphics[width=1\textwidth]{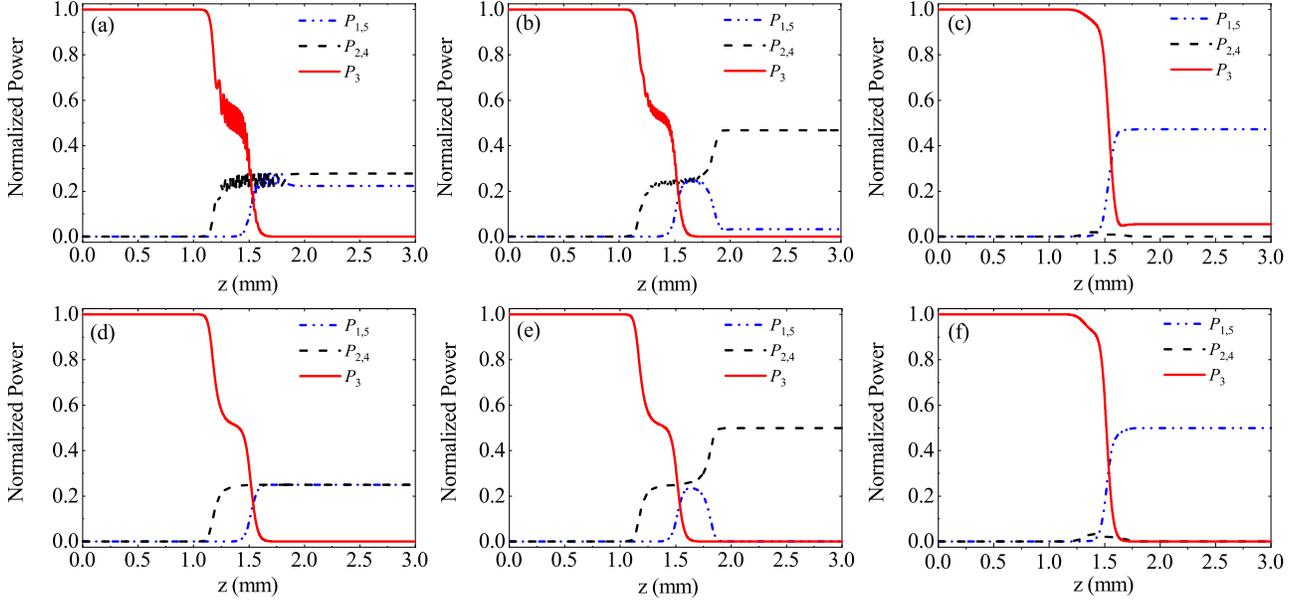}
\caption{Variable beam splitting between the middle WG and outer WGs for (a)-(c) without the STA  (upper row) and (d)-(e) with the STA (lower row). Parameters: (a) and (d): $P_{3}=0, P_{1,5}=P_{2,4}=0.25$ for $\Delta=-\frac{800}{3}z+400 (0<z\leq1.5) $ and $\Delta=0 (1.5<z\leq3)$. (b) and (e) : $P_{3}=0, P_{1,5}=0, P_{2,4}=0.5$ for $\Delta=-120z+180$. (c) and (f) : $P_{3}=0, P_{1,5}=0.5, P_{2,4}=0$ for $\Delta=1000$ with $z_{1}=1.6$, $z_{2}=1.4$ and $C_{0}=150$mm$^{-1}$ in all cases.}
\label{fig3}
\end{figure*}
Next, we analyze the more complex case where there is propagation constant mismatch between the even and odd WGs ($\Delta \neq 0$) and the WGs are lossless ($\Gamma=0$). Apparently, similar to the Case 1, the eigenvalues
\begin{equation}
\begin{aligned}
 E_{1}=&0,\\
 E_{2,4}=&\frac{\Delta}{2}\mp\frac{\sqrt{\Delta^2+4C^2_{1}}}{2},\\  E_{3,5}=&\frac{\Delta}{2}\mp\frac{\sqrt{\Delta^2+4C^2_{1}+8C^2_{2}}}{2},
\end{aligned}
\end{equation}
and we can also derive the instantaneous eigenstate corresponding to the eigenvalues $E_{4}$:
\begin{equation}
\begin{aligned}
|\Phi_{4}(z)\rangle=&\frac{\sqrt{2}\cos(\theta)\cos(\varphi)}{2\sqrt{1+\sin^2(\theta)}}|1\rangle-\frac{\sqrt{2}}{2}\sin(\varphi)|2\rangle,\\
&+\frac{\sqrt{2}\sin(\theta)\cos(\varphi)}{2\sqrt{1+\sin^2(\theta)}}|3\rangle-\frac{\sqrt{2}}{2}\sin(\varphi)|4\rangle,\\
&+\frac{\sqrt{2}\cos(\theta)\cos(\varphi)}{2\sqrt{1+\sin^2(\theta)}}|5\rangle,
\end{aligned}
\end{equation}
where the mixing angles are respectively defined as:
\begin{eqnarray}
\tan(\theta)=\frac{C_{2}}{C_{1}}, \tan(2\varphi)=\frac{2\sqrt{C^2_{1}+2C^2_{2}}}{\Delta}.
\end{eqnarray}
In the following, we utilize the STA protocol to design robust and shorter optical splitting devices. Indeed, the evolution is never perfectly adiabatic, and some nonadiabatic coupling are always present, which limits the efficiency of the STIRAP. In the STA scheme, an additional Hamiltonian $H_{cd}(z)=i\hbar\sum_{n}|\partial_{z}\Phi_{n}(z)\rangle\langle\Phi_{n}(z)|$, where $|\Phi_{n}(z)\rangle (n=1,2,3,4,5)$ is added to our original Hamiltonian $H(z)$ and the total Hamiltonian $H_{total}(z)=H(z)+H_{cd}(z)$. Now we can continue to predict the dynamics of light propagation in WGs which obeys the following adiabatic state $|\Phi_{4}(z)\rangle$. It is clear that if the coupling constants and the detuning between the WGs in the array are not the same, different distributions of the intensities can be achieved. Therefore, we can observe variable beam splitting in the WG structure under different choices of the detuning as shown in Fig. \ref{fig3}. The results display the evolution of population with different proportions, such as: $\mathbf{a}(z_{f})=\lbrack \frac{1}{2},\frac{1}{2},0,\frac{1}{2},\frac{1}{2}\rbrack^T$, $\mathbf{a}(z_{f})=\lbrack 0,\frac{\sqrt{2}}{2},0,\frac{\sqrt{2}}{2},0\rbrack^T$ and $\mathbf{a}(z_{f})=\lbrack \frac{\sqrt{2}}{2},0,0,0,\frac{\sqrt{2}}{2}\rbrack^T$. To demonstrate the advantages of STA protocol for accelerating adiabatic evolution, we set the same coupling coefficient and detuning as shown in Fig. \ref{fig3}. Moreover, as to the case of the large detuning between the propagation constants of the even and odd WGs in Fig. \ref{fig3} (c) and (f), the five-WG system can be reduced to an effective two-WG system, which is known as adiabatic elimination \cite{PhysRevA.97.023811}.
\subsection{Case 3: $\Delta\neq 0,\Gamma\neq 0$}
In this subsection, we consider the case where both the detuning and the dissipation exist. In this case, the system should be described by a non-Hermitian Hamiltonian, and the calculation becomes much more complicated. Nevertheless, we can still obtain the eigenvalues with analytical expressions written as
\begin{equation}
\begin{aligned}
E_{1}=&0\\
E_{2,4}=&\frac{\Delta-i\Gamma}{2}\pm\frac{\sqrt{(\Delta-i\Gamma)^2+4C^2_{1}}}{2}\\  E_{3,5}=&\frac{\Delta-i\Gamma}{2C_{1}}\mp\frac{\sqrt{(\Delta-i\Gamma)^2+4C^2_{1}+8C^2_{2}}}{2}
\text{.}
\end{aligned}
\label{eq14}
\end{equation}
There is, however, an essential distinction between the eigenvalues of the Hermitian and non-Hermitian method. While in the Hermitian case the norm of the eigenvalues remains real during the entire evolution, in the non-Hermitian (NH) case, as seen from Eq. (\ref{eq14}) , the eigenvalues can be real or complex. Likewise, we also follow the steps of Case 1 to choose the eigenstate with zero eigenvalue as the instantaneous eigenstate which given by
\begin{equation}
\begin{aligned}
|\Phi_{1}(z)\rangle=&\frac{\sin(\theta)}{\sqrt{1+\sin^2(\theta)}}|1\rangle-\frac{\cos(\theta)}{\sqrt{1+\sin^2(\theta)}}|3\rangle\\
&+\frac{\sin(\theta)}{\sqrt{1+\sin^2(\theta)}}|5\rangle
\text{,}
\end{aligned}
\end{equation}
where the mixing angles are respectively defined as:
\begin{eqnarray}
\tan(\theta)=\frac{C_{2}}{C_{1}}, \tan(2\varphi)=\frac{2\sqrt{C^2_{1}+2C^2_{2}}}{\Delta-i\Gamma}.
\end{eqnarray}
However, perfectly adiabatic evolution is hard to realize, and in a realistic physical circumstance the adiabatic criterion usually cannot be fulfilled, that is, complete beam splitting from WG3$\rightarrow $WG1 (WG5) does not occur due to the effect of the non-adiabatic coupling between adiabatical states. Therefore, the loss of light energy is very large by manipulating STIRAP, which is harmful to the fabrication of an integrated optical device. To overcome this detriment of STIRAP, we exploit NH-STA to make beam splitters more efficient and small in dimension. This method is performed by introducing a series of redesigned supplementary Hamiltonians to nullify the specified non-adiabatic couplings and the system is subjected to the total Hamiltonian $H_{total}(z)$ with the correction
\begin{eqnarray}
H_{cd}(z)=i\hbar\sum_{m \neq n}\sum \frac{\Pi_{m}|\dot{H}(z)|\Pi_{n}}{E_{n}(z)-E_{m}(z)},
\end{eqnarray}
where $\Pi_{m}=|\Phi_{m}(z)\rangle\langle {\Phi}_{m}(z)|$ and $\Pi_{n}=|\Phi_{n}(z)\rangle\langle {\Phi}_{n}(z)|$, $|\Phi_{m}(z)\rangle$ and $| {\Phi}_{n}(z)\rangle$ $(m,n=1,2,3\cdots,5)$ are defined as the eigenstates of $H(z)$ and $H^{\dag}(z)$ respectively.

We consider the input as $\mathbf{a}(z_{i})=\lbrack 0,0,1,0,0\rbrack^T$ again, for the following initial and final conditions in Eq. (\ref{eq10}) and the following final light field distributions among WG: $\mathbf{a}(z_{f})=\lbrack \frac{\sqrt{2}}{2},0,0,0,\frac{\sqrt{2}}{2}\rbrack^T$ (ideally).
\begin{figure}[htbp]
\centering
\includegraphics[width=0.45\textwidth]{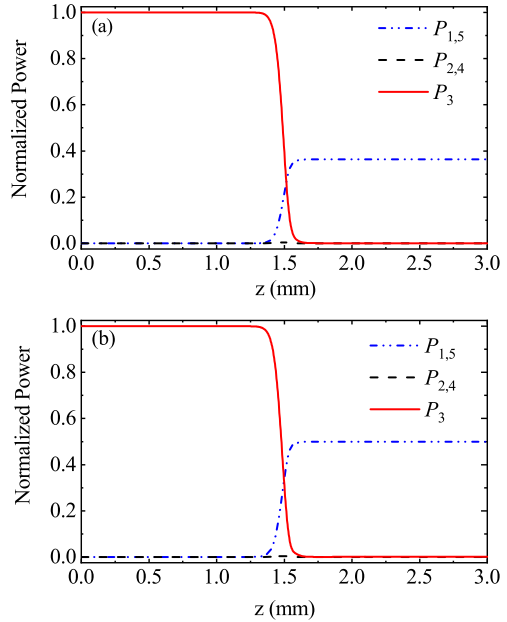}
\caption{Efficiency of beam splitting without the NH-STA (a) and (b) with the NH-STA. Parameters: $C_{0} =100$mm$^{-1}$, $\Delta=200$mm$^{-1}$, $\Gamma=200$mm$^{-1}$, $z_{1}=1.4$mm and $z_{2}=1.6$mm.}
\label{fig4}
\end{figure}
In Fig. \ref{fig4}, we compare the evolution of the populations and for the cases without/with the NH-STA. Through the results as without the NH-STA, we can conclude that if the WG2 and the WG4 are not populated during the evolution, the loss rate of these two WGs have a little effect on the evolution of the system. However, since the existence of detuning and dissipation violates the adiabatic condition, the system no longer completely follows the dark state. It should be noted here that, since the Hamiltonian is non-Hermitian that the loss has to be included, the norm of the state vector does not need to be conserved during the evolution. Contrastively,
it can be seen that in the case of NH-STA the power splitting strongly outperforms that without the NH-STA, which is robust against the propagation loss of outer WGs. In this way, the light is splitted equally between two outermost WGs, which are helpful to design robust WG splitters, while without the NH-STA we have only $P_{1,5}\approx0.36$ for the same values of $C_{0}$ and $z$. Furthermore, to compare the performance of the proposed NH-STA splitter with the original one, we show the contour plots of the light intensity at the end of the device at WG5 (WG1) as a function of $C_{0}$ and $\Gamma$ in Fig. \ref{fig5}. We note that the fidelity of the light transfer for the NH-STA coupler is robust against variations in both $C_{0}$ and $\Gamma$.
\begin{figure}[htbp]
\centering
\includegraphics[width=0.4\textwidth]{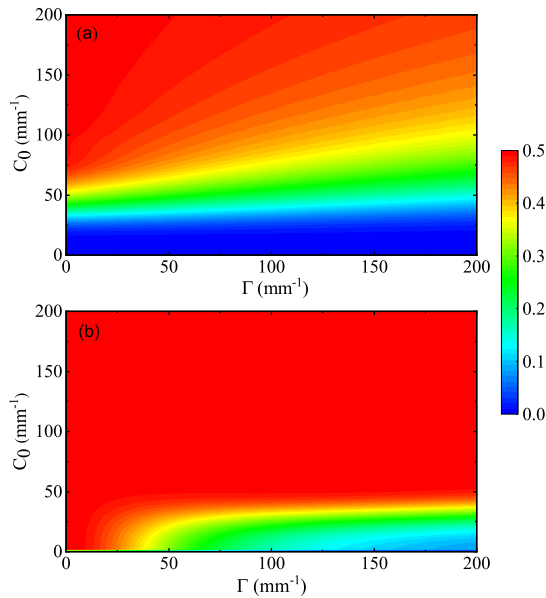}
\caption{Contour-color plot of the output power $P_{1,5}$ as obtained from an integration of Eq. (\ref{eq1}) as a function of $C_{0}$ and $\Gamma$ without the NH-STA (top) and with the NH-STA (bottom) with $\Delta=200$ mm$^{-1}$.}
\label{fig5}
\end{figure}
\subsection{Fabrication tolerances and performance comparison}
\begin{figure}[htbp]
\centering
\includegraphics[width=0.45\textwidth]{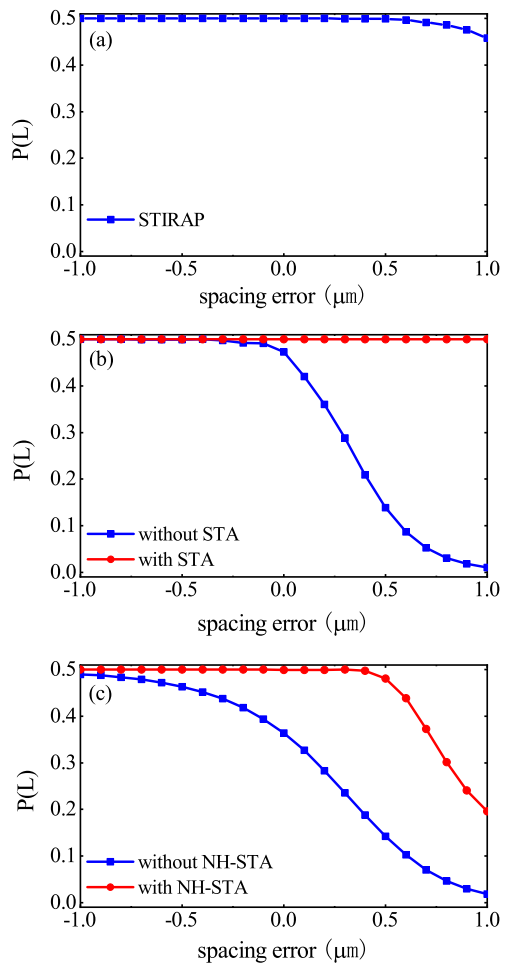}
\caption{Output power in WG1 (5) as a function of WG spacing error at $L=3$mm as calculated by the Eq. (\ref{eq1}). The STIRAP in Fig. \ref{fig2} (b) (top) , without the STA in Fig. \ref{fig3} (e) and with the STA in Fig. \ref{fig3} (f) (middle) and without the NH-STA in Fig. \ref{fig4} (a) and with the NH-STA in Fig. \ref{fig4} (b) (bottom).}
\label{fig6}
\end{figure}

In order to examine the fabrication tolerances of our beam splitters, we add a constant error to the WG spacing. For simplicity, we take our $1\times2$ beam splitters as an example to analyze the tolerance of three different types of beam splitters respectively. The simulation result is shown in Fig. \ref{fig6}. As clearly illustrated in Fig. \ref{fig6} (a), the normal STIRAP evolution without detuning and loss is insensitive to spacing error within a certain range. It is also demonstrated in Fig. \ref{fig6} (b) the spacing error has a little effect on the power splitting for $-1$ to $1$ $\mu$m for the optimized device via STA to eliminate the non-adiabatic coupling. The same behavior is depicted in Fig. \ref{fig6} (c), the power P(L) is close to $0.5$ in our NH-STA scheme for a spacing error as large as $-1$ to $0.25$ $\mu$m even with the power loss of the outer WGs. It can be seen that our $1\times2$ beam splitters described above present good robustness on the spacing error fluctuation.

On the other hand, we compare the parameters and performance of our works with other reported references in Table. \ref{tab:table1}. We note that there is an efficient distribution of power in the two output ports in Fig. \ref{fig2} (b) , Fig. \ref{fig3} (f) and Fig. \ref{fig4} (b) that is close to the ideal case where the transmission is $50\%$ at output 1 (5), with a total of $100\%$, these results show a good performance in terms of transmission. By comparison with two works \cite{Ciret:12,Lunghi:18} from the literature mentioned in the Table. \ref{tab:table1}, a high transmission with shorter device length are achieved in our works, which is beneficial to the experimental realization.
\begin{table*}[htbp]
\begin{center}
  \setlength{\abovecaptionskip}{0pt}
  \setlength{\belowcaptionskip}{10pt}
     \caption{$1\times2$ beam splitters parameters and transmission performances with other works
                 \protect \\ \qquad \qquad \qquad \qquad \qquad \qquad \qquad(Wavelength $\lambda= 1.55\mu$m).}
\label{tab:table1}
\setlength{\tabcolsep}{4mm}{
\begin{tabular}{clccccc}
\hline\hline
\multicolumn{2}{c}{\multirow{2}{*}{$1\times2$ beam splitters}}& WG &\multicolumn{4}{c}{Transmission(\%)} \\
\cline{4-7}
  & & length [$\mu$m] & Port 1 (5)& Port 2 (4)& Port 3  &Total\\
\hline
\multicolumn{2}{c}{Ref. \cite{Ciret:12}} & 4000  & 50 & 0 & 0 & 100\\
\hline
\multicolumn{2}{c}{Ref. \cite{Lunghi:18}} & 15000  & 49 & 0 & 0 & 98\\
\hline
\multicolumn{2}{c}{our works}\\
\hline
Case 1& STIRAP (Fig. \ref{fig2} (b)) & 3000  & 50 & 0 & 0 & 100\\
\hline
\multirow{2}{*}{Case 2}  & no STA (Fig. \ref{fig3} (c)) & 3000 & 47.29 & 0 & 5.42 & 100\\
\cline{2-7}
  &STA (Fig. \ref{fig3} (f)) & 3000  & 50 & 0 & 0 & 100\\
\hline
\multirow{2}{*}{Case 3} &no NH-STA (Fig. \ref{fig4} (a))&3000&36.37&0&0.16&72.9\\
\cline{2-7}
&NH-STA (Fig. \ref{fig4} (b))&3000&50&0&0&100\\
\hline\hline
\end{tabular}}
\end{center}
\end{table*}
\section{\label{sec:level4}Examples, spectral behavior}
\begin{figure}[htbp]
\centering
\includegraphics[width=0.45\textwidth]{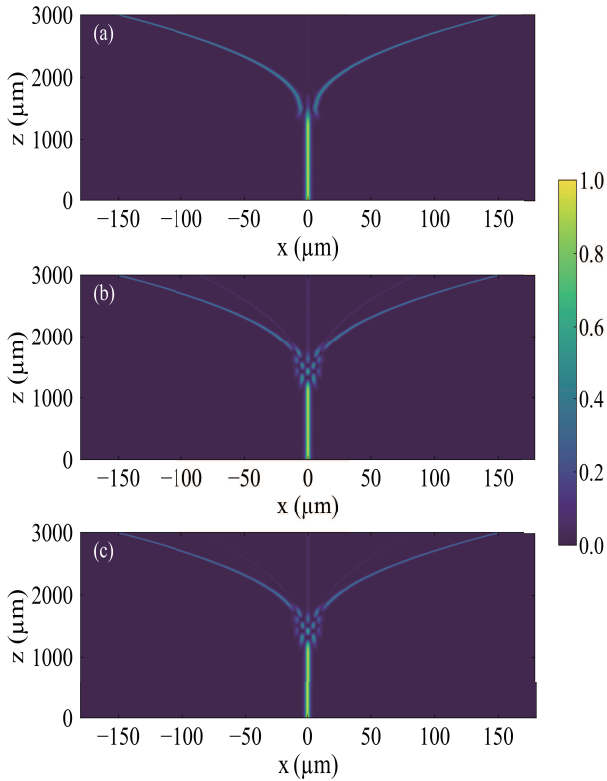}
\caption{Light wave evolution in a five-WG structure for $1\times2$ beam splitters calculated by the BPM method for light injected into the WG3 and output to the WG1 and WG5 for the Case 1-3, respectively. Here the minimum WG separation between the middle WG3 and the outer WG2 and WG4 $D^{(1)}_{0}= 3\mu$m$^{-1}$. During the optical simulations, for (b) , $\Delta n_{2}=\Delta n_{4}=0.012$ and the minimum WG separation between the outer WG2 and WG4 and the outerst WG1 and WG5 $D^{(2)}_{0}= 4.04\mu$m$^{-1}$ for Fig. \ref{fig3} (c), for (c) $\Delta n_{2}=\Delta n_{4}=0.0119$, $\Gamma=9.02$mm$^{-1}$ and $D^{(2)}_{0}= 3.96\mu$m$^{-1}$ for Fig. \ref{fig4} (a).}
\label{fig7}
\end{figure}

In the following, we give three concrete examples employing a split-step Fourier beam propagation method (BPM) \cite{kawano2004introduction}, which has been widely used to design optical beam splitters and predict the light evolution in WGs without further assumption. During the optical simulations, the substrate was fabricated with commercially available epoxy that $n_{s} = 1.5$ and 3$\mu$m thick SiO$_{2}$ on a Si wafer is used for the bottom cladding layer in our polymer WGs structure. The input wavelength $\lambda= 1.55$$\mu$m and the WG length $L=3$mm. In the nearly symmetric WG, the coupling coefficient $C$ and the WG separation $D$ are well fitted by the following exponential relation: $C=C_{0}\exp\lbrack-\gamma(D-D_{0})\rbrack$ \cite{Chen_2018}, where $C_{0}=1$mm$^{-1}$, $\gamma\simeq 1.409\mu$m$^{-1}$. We choose that all WGs have the same width W=$3\mu$m and $\Delta n=0.0116$. As to the space-dependent detuning $\Delta$, they can be approximately realized by linearly modifying the width difference $\delta$W  and the refractive index contrast difference $\delta n$ of the WGs  \cite{721079}. In Fig. \ref{fig7} (a)-(c), we demonstrate the light propagation in symmetric five WGs for $1\times2$ beam splitters.
The corresponding evolution of the standard STIRAP process in Fig. \ref{fig2} (b) for which five WGs have the same index profile and thus $\Delta=0$ is shown in Fig. \ref{fig7} (a). Since the adiabatic conditions of Eq. (\ref{eq9}) are satisfied, in this case, we can find that the outer WG2 and WG4 are almost not excited during the optical evolution. As seen in Fig. \ref{fig7} (b), when the WG2 and WG4 are detuned with odd ones, there is a small energy population on the WG2 and WG4 due to the presence of non-adiabatic coupling. And when the WG2 and WG4 are detuned with odd ones and are dissipative as in Fig. \ref{fig4} (a), we can also find that the output power in each of the outermost WGs is slightly less than $0.5$ at $z=L$ in Fig. \ref{fig7} (c). The compound WG systems can be fabricated experimentally by photolithography \cite{Kaino2002}, dry etching \cite{Ruan:04}, photobleaching \cite{Friebele:81} and other technologies. Among them, photolithography is widely used in WG fabrication. One can design lithographic masks according to the different beam splitters and then the relative distance between WGs along the propagation $z$ can be adjusted correspondingly.
\section{\label{sec:level5}Conclusions}
In summary, we have theoretically proposed various short and robust multiple beam splitters in Hermitian and non-Hermitian symmetric WGs. In Hermitian systems, we have analyzed resonant and phase mismatch coupling models. In resonant case, we have achieved an one-to-two beam splitting by STIRAP. In the propagation constants mismatch case, we have designed three proportions of light splitting couplers via STA technique to eliminate the non-adiabatic coupling in evolution process. In a non-Hermitian system with the even WGs being dissipative, we have also investigated the controllable multiple beam splitting by using NH-STA scheme. These results have showed that our devices are robust against the fluctuation of parameters, as the fidelity is still unity for a wide range of $C_{0}$ and $\Gamma$, and the even WGs being dissipative has little effect on the adiabatic beam splitting in the WGs systems. It is noted that the splitters of the arbitrary power radio between the middle WG and outer WGs can also be obtained by selecting appropriate coupling coefficient and detuning. The proposed device not only can reduce significantly the device length of the couplers, but also can keep an ultra-high fidelity compared with the regular WG couplers, which is interesting and valuable for studying the high-fidelity controllable multiple beam splitting in integrated optics.
\section*{Acknowledgments}
The work is supported by the National Natural Science Foundation of China (Grant No. 12075193).
\bibliography{refercence}
\end{document}